\documentstyle[12pt]{article}
\begin{document}

\newcommand{\preprintno}[1]
{\vspace{-2cm}{\normalsize\begin{flushright}#1\end{flushright}}\vspace{1cm}}

\title{\preprintno{{\bf SUSX-TH-99-017}\\ astro-ph/9910328}Carbon Burning in Supernovae and Evolving Physical Constants}
\author{Malcolm Fairbairn\thanks{E-mail address: mdsf@star.cpes.sussex.ac.uk}\\
        {\em Centre for Theoretical Physics / Astronomy Centre
, University of Sussex,} \\
        {\em Brighton BN1 9QH, U.K.}}
\date{October 1999}
\maketitle

\begin{abstract}
The C$^{12}\{$C$^{12}$,$\alpha\}$Ne$^{20}$ fusion reaction rate is shown to be extremely sensitive to variations in the strong coupling constant, $\alpha_{s}$.  Connection with $\alpha_{s}$ is established by using the one boson exchange potential for the nuclear forces and then relating the meson masses and coupling constants of this model to QCD.  The implications of a cosmological evolution of these constants on carbon burning in Type 1a Supernovae are discussed. 
\end{abstract}

\section{Introduction}

Type 1a Supernovae (henceforth SNe1a) are thought to result from the ignition of a degenerate Carbon/Oxygen stellar core.  This occurs when the mass of the core increases until the energy released via C$^{12}+$C$^{12}$ fusion can no longer be radiated away by neutrinos.  SNe1a explosions are badly understood and it is not clear (for example) at what mass the core begins runaway burning, where in the core burning is instigated and what kinds of progenitor systems lead to SNe1a's $\cite{waw94,nom99}$.  At low redshifts there is a strong relationship between SNe1a luminosity rise times and peak luminosities, rendering these objects potentially useful standard candles for cosmology. 

The low apparent luminosity of high redshift (0.3 $<$ z $<$ 1.0) supernovae can be interpreted as evidence for a FRW universe with a cosmological constant of $\Lambda \approx 0.7$ or as SNe1a evolution.  It has been suggested that the peak luminosity of a SNe1a is higher if there is a lower Carbon/Oxygen ratio in the core $\cite{hof98}$.  However, higher mass stars which evolve quickly and should therefore dominate the SNe1A population at high redshifts have lower Carbon abundance cores, creating a problem for this evolution argument.  At the same time, recent studies suggest that the rise times of high redshift SNe1a's are shorter than the rise times of local SNe1a objects, indicating a possible evolutionary effect $\cite{riess99}$.  The astrophysical analysis assumes that the nature of fundamental physical interactions have remained the same throughout the history of the Universe which, until tested rigorously, may not be the case.
   
Over the past 20 years it has become clear that theories involving the presence of higher dimensions can provide a framework in which the four forces of nature may be described by a unified theory $\cite{polchinski}$.  The fine structure constants in these theories are related to the size of the higher dimensions so it is neccesary to provide a mechanism to stabilise these sizes.  As it is still not entirely clear what this mechanism may be, any observational evidence for the evolution of fine structure constants over cosmological time scales might shed light on the problem.  At the same time, changing fine structure constants would lead to changes in the behaviour of supernovae.  Most physical processes would be affected including nuclear fusion cross sections and energy yields so the evolution and luminosity of the explosion would certainly be different.    
In $\cite{amend99}$ the effect of the time variability of the gravitational constant on the Chandrasekhar mass and the background cosmology were calculated and used to reconcile the supernova observations with a $\Lambda=0$ open universe.  In this article we investigate the effect of changing electromagnetic and strong fine structure constants on the carbon fusion responsible for SNe1a explosions.  As $i$) it is not clear as to whether SNe1a explosions occur at the Chandrasekhar mass $\cite{nom99}$ and $ii$) it is not clear how sensitive explosion luminosities are to the mass of the progenitor $\cite{waw}$, we will be less ambitious than the authors of $\cite{amend99}$ and simply calculate the effect of changing $\alpha_{em}$ and $\alpha_{s}$ on the carbon fusion cross section and energy yields (Q values).  

\subsection{Existing Limits on Variations in $\alpha_{em}$ and $\alpha_{s}$}
Webb et al. $\cite{webb}$ placed constraints on the fractional variation of $\alpha_{em}$ as being about 1 part in 10$^5$ at a redshift of z=1.  The variation of $\alpha_{s}$ is more subtle as, because of well known QCD renormalisation effects on the strong coupling constant, $\alpha_{s}$ becomes considerably larger at low energies.  The 3 colour charge one-loop renormalization group equation for the QCD coupling constant is given by

\begin{equation}
{\frac{2\pi}{\alpha_{s}(\rho)}}={{\frac{2\pi}{\alpha_{s}(\rho_{c})}}}+11ln\left(\frac{\rho_{c}}{\rho}\right)
\end{equation}
where $\rho$ has dimension of inverse mass or length and $\rho_{c}$ corresponds to some energy scale at which $\alpha_{s}$ is known.  This energy scale has to be within the regime of the three light quarks, up down and strange, in order that the above relation holds (at higher energies, virtual pairs of charm, bottom and top quarks will lead to corrections to the renormalisation group equation).  To say what $\alpha_{em}$ means today is straight forward, but the growing value of $\alpha_{s}$ at low energies means that any QCD physics observed today occurs within highly bound systems such as the proton.  Also, as one can see by examining the renormalization equation, small changes in $\alpha_{s}$ at high energies correspond to very large changes in $\alpha_{s}$ at low energies.  Consequently, not only is it necessary to decide how much one is permitted to change $\alpha_{s}$ but also at what energy scale this constraint is imposed.  A very strict limit on the varying of $\alpha_{s}$ has been provided by considering capture cross sections for slow neutrons $\cite{sky76}$.  If one saturates this limit and assumes the rate of change is constant, a 2$\%$ limit on the fractional variation of $\alpha_{s}$ is obtained for a time 10$^{10}$ years ago.  Although these capture cross sections were observed for thermal energies much lower than the mass of the proton, the absorption will depend on the wave function of the absorbed neutron within the nucleus.  This wave function will involve a potential derived from the interactions of nuclei at an energy scale of the order of the nuclear size and hence the proton mass.  This is the energy scale at which $\alpha_{s}$ has been varied in this study.

\section{The C$^{12}+$C$^{12}$ Fusion Reaction}

\subsection{Fusion Cross Section Theory}

The reaction rate for the fusion of two nuclei is given by $\cite{rnr}$
\begin{equation}
\langle{\sigma}v\rangle=\left({\frac{8}{\pi\mu}}\right)^{1/2}{\frac{1}{(kT)^{3/2}}}\int_{0}^{\infty}\sigma(E)E exp\left(-{\frac{E}{kT}}\right)dE
\end{equation}
where $\mu$ is the reduced mass, T the temperature, E is energy, $\sigma(E)$ is the reaction cross section, k is Boltzmann's constant and v is the velocity of particles in the medium.  The cross section can be written
\begin{equation}
\sigma(E)={\frac{1}{E}}exp(-2{\pi}{\frac{Z_{1}Z_{2}e^{2}}{{\hbar}v}})S(E)
\end{equation} where $Z_{1}$ is the number of protons in the target nucleus and $Z_{2}$ refers to the number of protons in the incident nucleus.  S(E) parametrises the resonant structure due to different channels being favoured at different energies.  Theoretical estimates of $S(E)$ for C$^{12}+$C$^{12}$ fusion have been particularly controversial as early laboratory experiments suggested the presence of pronounced resonances at astrophysical energies (up to a few hundred keV) $\cite{fowler}$.  These were thought to be due to the phenomenon of absorption under the barrier where the two nuclei form an intermediate nuclear molecular state.  Several approaches to modelling this interaction were investigated, none of which predicted the observed resonant structure entirely successfully $\cite{mich72}$$\cite{mich73}$$\cite{arn76}$$\cite{desc89}$.  However, in $\cite{beck}$, it was shown that carbon fusion was modelled acceptably by the WKB approximation where the cross section is written $\cite{stead86}$
\begin{equation}
\sigma(E)=\pi\bar{\lambda}^{2}(E)\sum^{\infty}_{l=0}(2l+1)T_{l}(E)\\
\end{equation}
$\lambda$ being the reduced deBroglie wavelength of the incident nucleus and $\ell$ being the angular momentum number. The transmission coefficients $T_{l}(E)$ are written
\begin{equation}
T_{l}(E)=[1+exp(S_{l})]^{-1}
\end{equation}
where $S_{l}$ is the classical action evaluated between the inner and outer classical turning points, R$_{\textnormal{i}}$ and R$_{\textnormal{o}}$.
\begin{equation}
S_{l}=2\sqrt{{\frac{2\mu}{\hbar^{2}}}{\int^{R_{o}}_{R_{i}}[V_{l}(r)-E]dr}}
\end{equation}

Here, $\mu$ is the reduced mass and the potential V$_{l}$(r) is given by 
\begin{equation}
V_{l}(r)=V_{coul}(r)+V_{nuc}(r)+{\frac{l(l+1)\hbar^{2}}{2{\mu}r^{2}}}.
\end{equation}
The Gamow Peak energy, $E_{Gam}$, is the favoured energy for fusion in thermalised systems and corresponds to the energy at which the product of the Maxwell-Boltzmann distribution and the barrier penetration probability grows highest.  Evaluation of the dominant $\ell$=0 channel cross section at $E_{Gam}$ will give a reasonable estimate of the change in the astrophysical fusion rate as the nuclear and Coulombic potentials change.  The Gamow peak energy is given by $\cite{rnr}$
\begin{equation}
E_{Gam}=\left(\sqrt{\frac{\mu}{2}}{\frac{{\pi}e^{2}Z_{1}Z_{2}kT}{\hbar}}\right)^{2/3}
\end{equation}
\subsection{Inter-Nucleus Potential Model}
In order to evaluate the cross section, the potential due to both Coulombic and nuclear forces between the two carbon nuclei must first be calculated.  This was done by modelling each carbon nucleus as a solid sphere of constant density with radius
\begin{equation}
R=r_{\textnormal{proton}}A^{1/3}=2.28984r_{\textnormal{proton}}.
\end{equation}
Experimental data shows that this saturated density ansatz is a good approximation, although there are models thought to be more realistic where density tails off exponentially from the surface $\cite{arn76}$.  Equation (9) shows that one can calculate how the radius of any nucleus changes when constants are altered given that we know how the proton radius reacts.  The success of the saturated density model also indicates that the nucleus structure is dominated by short range effects and consequently by the strong force as opposed to electromagnetism.    

Next, we assume that the twelve units of nuclear charge (Baryon Number) and six units of Coulombic charge are evenly distributed throughout the nucleus.  Then it is possible to do a numerical integration by approximating the two nuclei as spherical sections of a three dimensional grid.  The potential energy at each individual site due to the nuclear charge located at one of the sites on the other nucleus is given by the single nucleon-nucleon one boson exchange potential with appropriate weighting.  Because spins of individual nuclei are neglected in this simple model, the nucleon-nucleon potential is vastly simplified as we shall see in the next section.\par
The cross section was calculated with all parameters fixed as observed in nature, then $\alpha_{em}$ and $\alpha_{s}$ were altered independently.  When $\alpha_{em}$ is altered, the Coulombic component of the inter-nuclear potential and the binding energies of the nuclei are altered.  When $\alpha_{s}$ is altered at some energy, the energy scale at which quark confinement occurs (approximately when $\alpha_{s}(\rho)$ in (1) goes to one) also changes.  This tells us how the proton radius and mass, and therefore the nuclear radius and mass, change and the change in the masses of the mesons in the boson exchange potential also can be calculated as we shall see later.  Again, the binding energies and consequently the energy liberated by the reaction are changed by varying $\alpha_{s}$.  These effects were included in the simulations.

\section{The One Boson Exchange Potential}

In order to calculate how the nuclear force between two C$^{12}$ nuclei will vary with QCD parameters it is first necessary to consider the single nucleon-nucleon potential.  This is described quite successfully by the one boson exchange potential (OBEP) $\cite{mach}$. 
The most significant interactions of the four meson fields that play a role in the OBEP are the pseudoscalar interactions of the isovector $\pi$, the attractive exchange of the scalar boson $\sigma$, the repulsive vector interactions of the isoscalar $\omega$ and the tensor interactions  of the isovector $\rho$ $\cite{mach}$.  The tensor potential has an average value of zero upon integration over all angles of incidence $\cite{gam}$ so the $\rho$ potential will be neglected as we will not be using a complicated C$^{12}$ model.  Further simplification results from the fact that the pseudoscalar potential describing single $\pi$ exchange contains a factor ${\bf{\sigma}}_{1}\cdot{\bf{\sigma}}_{2}$ taking into account the relative spins of the two spin 1/2 nucleons.  The value of this product is either -3 for the singlet (degeneracy 1) spin state or +1 for the triplet (degeneracy 3) spin states $\cite{blatt}$ so if we make no assumptions about the relative orientation of the spins of the nuclei in the model, the average force due to single $\pi$ exchange vanishes.
We are subsequently left with the scalar $\sigma$ and the repulsive vector $\omega$ potentials which can now be written as $\cite{mach}$
\begin{eqnarray}
V_{\sigma}(m_{\sigma},r)&=&-{\frac{g_{{\sigma}NN}^{2}}{4\pi}}{\frac{e^{-m_{\sigma}r}}{r}}\\
V_{\omega}(m_{\omega},r)&=&\left\{{\frac{g_{{\omega}NN}^{2}}{4\pi}}m_{\omega}\left[\left(1+{\frac{1}{2}}{\left({\frac{m_{\omega}}{m_{n}}}\right)}^{2}\right)-{\frac{3}{4M^{2}}}\left( {\frac{m_{\omega}^{2}}{r}}+{\frac{2m_{\omega}}{r^2}}+{\frac{2}{r^3}}\right)\right]\right.\nonumber\\
&+&\left.{\frac{1}{2}}{\frac{g_{{\omega}NN}f_{\omega}}{4\pi}}m_{\omega}\left({\frac{m_{\omega}}{m_{n}}}\right)^{2}\right\}{\frac{e^{-m_{\omega}r}}{m_{\omega}r}}.  
\end{eqnarray}
Here, $m_{\omega}, m_{\sigma}$ are the masses of the omega and sigma mesons, $g_{{\sigma}NN}, g_{{\omega}NN}$ are the respective meson-nucleon-nucleon coupling constants, $f_{\omega}$ is the omega meson decay constant and $m_{n}$ is the nucleon mass.
The next step is to relate the parameters of this potential to the strong fine structure constant.

\section{Meson Masses and Coupling Constants from QCD}

As we shall see later in this section, the QCD vacuum is characterised by quark and gluon condensates.  In order to make contact between the OBEP and $\alpha_{s}$ it is necessary to derive the meson masses and coupling constants in terms of these condensates.

\subsection{$\sigma$ Meson Mass and Coupling Constant}

QCD sum rules were used by Ioffe to obtain an approximate expression for the nucleon mass in terms of the quark condensate $\cite{ioffe81}$
\begin{equation}
m_{n}={\frac{-2(2\pi)^{2}{\langle}q\bar{q}{\rangle}}{M^{2}}}
\end{equation}
where m$_{n}$ is the mass of the nucleon and M is the Borel mass.  This is a characteristic mass scale of the system introduced to regularise the series of contributions from higher dimensional operators which enter into light quark calculations $\cite{shifbig}$.  By assuming the Borel mass is the nucleon mass, one can write the approximate equation
\begin{equation}
m_{n}=[-2(2\pi)^{2}{\langle}q\bar{q}\rangle]^{1/3}
\end{equation}  
However, nucleon mass can be explained more satisfactorily by the Nambu Jona-Lasinio (NJL) model $\cite{njl}$$\cite{hats94}$. In this model, the nucleon acquires mass by interacting with the quark condensates of the vacuum via a scalar field.  Brown and Adami use an argument based upon the dependence of both sides of Ioffe's expression on N$_{c}$, the number of colour charges in the theory (3 for the SU(3) of QCD) to identify the scalar field of the NJL model with the sigma meson.  This results in the identification of the Borel mass not with the mass of the nucleon but with the sigma mass $\cite{brown91}$,${\cite{adami91}}$.  Using a constituent quark model for this process, they were able to write
\begin{equation}
m_{n}=-{\frac{2}{3}}{\frac{g_{{\sigma}NN}^{2}}{m_{\sigma}^{2}}}{\langle}q\bar{q}{\rangle}.
\end{equation}
Comparison of the above equations completely defines the sigma meson mass in terms of the quark condensate and the nucleon mass
\begin{equation}
m_{\sigma}=g_{{\sigma}NN}\left({\frac{2}{3}}{\frac{{\langle}q\bar{q}\rangle}{m_{n}}}\right)^{1/2}
\end{equation}
and sets the coupling constant value 
\begin{equation}
g_{{\sigma}NN}^{2}=12\pi^{2}.
\end{equation}
Consequently, the change in the proton radius (inverse mass) will have to be taken into account when calculating the change in the $\sigma$ meson mass. 

\subsection{$\omega$ Meson Mass and Coupling Constant}

The mass of the $\omega$ meson was predicted very accurately as long ago as 1977 when Chan considered the quark-loop expansion of the vector meson form factor $\cite{chan}$.  The result is simply
\begin{equation}
m_{\omega}=m_{\rho}=2\sqrt{2}{\pi}f_{\pi}\left({\frac{3}{N_{c}}}\right)^{1/2}.
\end{equation}  
The degeneracy between the $\omega$ and $\rho$ meson masses is broken by the fact that each has different total isospin and therefore mixes with virtual meson pairs differently (see e.g. $\cite{pich99}$).  However the approximation of Chan is still in excellent agreement with observation (about two percent $\cite{chan}$,$\cite{mach}$).  Now the omega coupling constant is related to the omega mass and the pion decay constant, $f_{\pi}$ by $\cite{mei88}$
\begin{equation}
g_{{\omega}NN}=\left({\frac{N_{c}}{2\sqrt{2}}}\right){\frac{m_{\omega}}{f_{\pi}}}=\pi\sqrt{3N_{c}}
\end{equation}  
determining all the parameters of the model in terms of the pion decay constant and the quark condensate.

\subsection{The QCD Vacuum}

A classical solution of the equation of motion in a Euclidean space-time sometimes represents quantum tunnelling through a barrier in a Minkowski space-time $\cite{hooft76}$.  Consequently, the discovery in the 1970's of an exact finite action Euclidean solution to the classical Yang-Mills equations lead to the realization that the QCD vacuum is highly non-trivial.  At short distances where the effective coupling $(\propto\sqrt{\alpha_{s}})$ is small, the dominant contributions to the path integral characterising the true vacuum are from tunnelling transitions between an infinite number of degenerate vacua $\cite{cdg78}$.  The vacuum is therefore determined at these distances by instanton effects rather than perturbation theory.  The way the density of instantons changes at different length scales $\rho$ in the colour SU(3) theory is given by the weighting function $\cite{cdg78}$
\begin{equation}
D_{SU(3)}(\rho)=(0.10)\left(\frac{2\pi}{\alpha_{s}(\rho)}\right)^{6}exp\left(-\frac{2\pi}{\alpha_{s}(\rho)}\right).
\end{equation}    
As one goes to larger distances, the effective coupling rises and so one must consider quantum corrections to the vacuum.  However, even though individual tunneling amplitudes are small,  the rapidly increasing number of instantons at larger radii means that tunneling contributions are still more important than loop corrections.  The qualitative explanation for confinement in this model is that small instantons and anti-instantons act to screen larger instantons, the effect of which are only felt at distances corresponding to their size, and the effective coupling rises rapidly at larger distances - hence confining any particle with colour charge.  Quarks carry colour charge so the distance at which the coupling constant becomes large is directly associated with the proton radius $\cite{ssv78}$.\par 
The fact that the classical description forms the bulk of the vacuum structure at small radii allows the use of the dilute instanton gas approximation to determine the self interaction of gluons and consequently an expression for the gluon condensate $\cite{shifbig}$
\begin{equation}
\left\langle{\frac{\alpha_s}{\pi}}G^{\mu\nu}G_{\mu\nu}\right\rangle=16\int_{0}^{\rho_{c}}{\frac{d\rho}{\rho^{5}}}D_{SU(3)}(\rho)
\end{equation}
where $G_{\mu\nu}$ is the gluon tensor and $\rho_{c}$ is an infrared cut-off corresponding to the length scale at which the effective coupling goes to one.

In the above picture, the QCD vacuum energy is obtained from the trace of the stress energy tensor which receives contributions from the quark and gluon condensates and any non-zero bare quark masses.  The QCD vacuum structure can also be obtained by considering the effective potential for composite operators $\cite{cornwall}$.  In this scheme, the vacuum energy will be a function of the bare quark masses and the quark condensate alone.  By comparing these two methods it is possible to relate the gluon and quark condensates.  In this work, we do calculations at the chiral limit where the up and down quarks have no bare masses.  Quark masses are hence entirely dynamically generated by vacuum screening effects.  At this limit, the relationship between the quark and gluon tensor has been recently established $\cite{gorbar}$  
\begin{equation}
\left\langle\bar{q}q\right\rangle\approx-\left[{\frac{1}{\epsilon}}\left\langle{\frac{\alpha_{s}}{\pi}}G^{\mu\nu}G_{\mu\nu}\right\rangle\right]^{3/4}
\end{equation}
where $\epsilon^{-1}$ can be shown to be $\cite{gorbar}$
\begin{equation}
{\frac{1}{\epsilon}}={\frac{16\pi{^2}}{N_{c}(2ln2-1)}}\left[{\frac{\beta(\alpha_{s})}{16\sqrt{\pi\alpha_{s}^{3}}}}+{\frac{17N_{c}ln2}{96\pi}}\right]
\end{equation}
and $\beta(\alpha_{s})$ is the perturbative 2 loop beta function
\begin{equation}
{\frac{\beta(\alpha_{s})}{16\sqrt{\pi\alpha_{s}^{3}}}}=-{\frac{1}{96\pi}}\left[(11N_{c}-2n_{f})+{\frac{1}{4}}{\frac{\alpha_{s}}{\pi}}\left(34N_{c}^{2}-10N_{c}n_{f}-3n_{f}{\frac{N_{c}^{2}-1}{N_{c}}}\right)\right]
\end{equation}
$N_{c}$ and $n_{f}$ being number of colours and flavours respectively.

Equation (20) can be written as an expansion in $\alpha_{s}(\rho_{c})$ $\cite{ssv78}$
\begin{equation}
\left\langle{\frac{\alpha_{s}}{\pi}}G^{\mu\nu}G_{\mu\nu}\right\rangle={\frac{64}{77}}{\frac{0.06}{\rho_{c}^{4}}}\left({\frac{2\pi}{\alpha_{s}(\rho_{c})}}\right)^{6}\left[1+\sum_{k=1}^{6}\left({\frac{11\alpha_{s}(\rho_{c})}{14\pi}}\right)^{k}{\frac{6!}{(6-k)!}}\right]exp\left(-{\frac{2\pi}{\alpha_{s}(\rho_{c})}}\right)
\end{equation}  
the dominant contribution to the expansion coming from the unity term inside the square brackets.  The value of the gluon condensate can be fixed by observation of the mass splitting of the ground states of J/$\psi$ and $\eta_{c}$ mesons $\cite{ssv78}$.  By using the observed value of 0.012 GeV$^4$ and setting $\alpha_{s}(M_{Prot})=1$ one can obtain a value of $\rho_{c}\approx1/125 MeV$, within a factor of 8 of the proton radius.  We then recycle this factor 8 to replace $\rho_{c}$ with the inverse proton mass in equation (24).  As we know how the proton mass varies with the strength of the strong force, we assume that $\rho_{c}$ varies in the same way.

The dynamically generated quark mass in the chiral limit at low momentum scales is given by $\cite{gorbar}$
\begin{equation}
m_{dyn}=\langle-q\bar{q}\rangle^{1/3}
\end{equation}
then the pion decay constant can be written $\cite{reinders}$
\begin{equation}
f_{\pi}={\frac{2^{1/6}}{\sqrt{2\pi}3^{1/4}}}m_{dyn}=0.34025\langle-q\bar{q}\rangle^{1/3}
\end{equation}
which determines all of the meson parameters in terms of quantities from QCD. 

\section{Binding Energies and Q-Values}
Here we make the assumption that despite the fact that shell closure effects render the liquid drop model inaccurate for nucleons with A$<$20 $\cite{blatt}$, the model can give an indication of the dependance of the nuclear binding energy on $\alpha_{em}$ and $\alpha_{s}$. The Semiemperical Mass formula is given by
\begin{equation}
M(A)=ZM_{prot}+NM_{neut}-a_{1}A+a_{2}A^{2/3}+a_{3}{\frac{Z^{2}}{a^{1/3}}}+a_{4}{\frac{(Z-N)^{2}}{A}}+\delta(A)
\end{equation}
where the terms with coefficients $a_{1}$ - $a_{4}$ are the volume, surface, Coulomb and asymmetry terms respectively.  The dominant channel for Carbon Burning is C$^{12}\{$C$^{12}$,$\alpha\}$Ne$^{20}$ (C$^{12}+$C$^{12}\Rightarrow$Ne$^{20}+\alpha$).  The volume contribution to the Carbon-Carbon system is the same as that for the $\alpha$-Neon system and consequently cancels out of the Q-value calculation.  The pairing term $\delta(A)$ gives a positive contribution to the mass for nuclei with odd numbers of both protons and neutrons (odd-odd), a negative contribution of the same magnitude for even-even nuclei and zero contribution for odd-even nuclei.  As Carbon nuclei, $\alpha$ particles and Neon nuclei are all even-even, the final state has the same ammount of pairing energy as the inital state and we do not need to calculate the dependance on r$_{prot}$ of the binding energies for this reaction.  The assymetry term can be neglected for all three nuclei, leaving us with the surface, Coulomb and pairing terms.  At this stage we still label the strength of electromagnetism directly by $\alpha_{em}$ but the contrived root from $\alpha_{s}(\rho)$ to observable physics makes it easier to signify the strength of the strong force by $r_{prot}$.  The empirical prefactors for the surface and Coulomb terms can be estimated quite well by using a simple Fermi gas model for the nucleus.     

This model uses the fact that only two neutrons and two protons can be in the same place with the same momentum to relate the density of nucleons to the highest occupied momentum state, the Fermi momentum k$_{F}$.
\begin{equation}
{\frac{2}{3\pi^{2}}}k_{F}^{2}=\rho={\frac{3m_{nuc}}{4\pi{r_{nuc}^{3}}}}\approx{\frac{3m_{prot}}{4\pi{r_{prot}^{3}}}}
\end{equation}
Since k$_{F}$ is proportional to the constant nuclear density and the nucleon mass is the inverse of the nuclear radius, k$_{F}\propto$r$_{prot}^{-4/3}$. The Fermi energy is therefore given by $\epsilon_{F}\propto$r$_{prot}^{-5/3}$.  In $\cite{deshalit}$, it is shown that the surface term coefficient as calculated by Fermi Gas considerations is only about 3$\%$ different from the empirical value.  The surface coefficient is obtained from the model by remembering to neglect any unphysical momentum states which occur when one or more components of the momentum vector are zero.  The result is
\begin{equation}
a_{2}={\frac{9\pi\epsilon_{F}}{40k_{F}r_{prot}}}{\propto}r_{prot}^{-4/3}.  
\end{equation}
Similarly, the Coulombic part of the energy can be derived from the Fermi gas model, although not as accurately as the surface term.  The full expression for the normal and altered terms is then
\begin{equation}
E_{Coul}={\frac{3}{5}}{\frac{Z^{2}e^{2}}{R}}-{\frac{9\pi}{4}}{\frac{Z^{2}e^{2}}{Vk_{F}^{2}}}={\frac{3}{5}}{\frac{Z^{2}e^{2}}{R}}{\frac{\alpha_{em}^{\prime}r_{prot}}{\alpha_{em}r_{prot}^{\prime}}}-{\frac{9\pi}{4}}{\frac{Z^{2}e^{2}}{Vk_{F}^{2}}}\left({\frac{r_{prot}}{r_{prot}^{\prime}}}\right)^{1/3}{\frac{\alpha_{em}^{\prime}}{\alpha_{em}}}. 
\end{equation}

The Q value for the C$^{12}\{$C$^{12}$,$\alpha\}$Ne$^{20}$ reaction using the semi empirical mass formula is hence
\begin{equation}
Q=0.595a_{2(surface)}-7.911a_{3(Coulomb)} 
\end{equation} 
This yields a value of 5.37 MeV, about 15$\%$ out from the observed value of 4.62 MeV $\cite{rnr}$ but we were expecting a discrepancy, as the liquid drop model is not entirely reliable for A$<$20.  We shall continue using this approximate method in the hope it will give us an indication of the magnitude of the effect of changing fine structure constants. 

\section{Results and Discussion}
When $\alpha_{em}$ is varied in the equations set out in section 2, the Carbon fusion cross section changes significantly more.  Saturating the bounds set by Webb of 1x10$^{-5}$ at a redshift of z=1, it is found that the fractional cross section change is 5.4x10$^{-4}$, a factor of 50 greater fractional change than $\alpha_{em}$.  However this is still very small and will make probably make very little difference to overall luminosities as we shall see below.

Using the equations in section 4 it was found that for a 2$\%$ increase/decrease in $\alpha_{s}$ at the proton mass, the meson masses increase/decrease by about 2.5$\%$ (both are $\propto\langle{q}\bar{q}\rangle^{1/3}$).  The same change in $\alpha_{s}$ resulted in an approximate 1.2$\%$ change in the radius of the proton, the radius decreasing as the coupling was increased.  Next, the effect of the changing potentials on the the l=0 cross section at the Gamow peak corresponding to a temperature of $1{\times}10^{9}$K were calculated ($1{\times}10^{9}$K is the typical temperature quoted for the time when the supernova is making the transition from quiescent to explosive burning $\cite{waw86}$).

Increasing $\alpha_{s}(M_{Prot})$ by 2$\%$ was found to decrease the cross section by 51$\%$ and decreasing $\alpha_{s}(M_{Prot})$ by 2$\%$ increases the cross section by 115$\%$, so both ways change the cross section by about a factor of two. The Q-value of the reaction also rises as $\alpha_{s}$ goes down, with ${\triangle}Q/Q\approx0.04$.

The fact that a 2$\%$ change in $\alpha_{s}$ yields can halve or double the fusion cross section is remarkable and could lead to interesting effects.  As an example, one might expect that if $\alpha_{s}$ had been 2$\%$ smaller at some time since the big bang, Carbon burning would take place much more readily, meaning that degenerate cores would explode at lower temperatures - hence lower masses.  A lower mass core would probably result in a lower luminosity explosion $\cite{waw86}$ and we would have a new interpretation of the apparent low luminosity of high redshift SNe1a's.  Unfortunately, things are not this simple because of the extreme temperature sensitivity of the Carbon fusion reaction.  

The apparently large change in the fusion cross section would be compensated for by a very small change in temperature.  One way of quantifying this effect is by taking the fusion cross section and multiplying it by the Q value to obtain a quantity signifying the energy yield for the three different values of $\alpha_{s}$.  It turns out that a 6.7$\%$ increase in temperature would compensate for a $2\%$ decrease in $\alpha_{s}$.  In a SNe1a this would certainly result in a change in the rate at which the transition from quiescent to explosive burning occurs and the peak luminosity would also change.  However, the exact way this would effect the relationship between rise-time and peak luminosity is difficult to quantify without a more complete understanding of SNe1a's.   

Models of Type 1a Supernovae are extremely complex, using detailed nuclear reaction schemes that take into account many different cross sections.  The fact that these models are still not entirely succesful after thirty years of endeavour show how complicated the mechanisms occuring in SNe1a's are.  It seems there are many other effects which may be responsible for the apparent evolution of SNe1a luminosity which do not involve speculative physics.  However, if there is reason to believe in a cosmological evolution of $\alpha_{s}$, SNe1a's may be a good place to start looking.  

\section*{Acknowledgments}
The author has had valuable discussions with the following people without which this work would have been impossible:  D.Bailin, J.D.Barrow, E.J.Copeland, T.Dent, J.P.Elliott and Sir D.Wilkinson.  This work is funded by PPARC.

\end{document}